\documentclass[twoside]{dis04}
\usepackage{amsmath}
\def\runauthor{M. Hirai, S. Kumano, N. Saito}
\def\shorttitle{AAC analysis of polarised parton distributions
                with uncertainties}
\begin{document}
%%%%%%%%%%%%%%%%%%%%%%%%%%%%%%%%%%%%%%%%%%%%%%%%%%%%%%%%%%%%%%%%%%%%%%%%%%%%%%%%

\title{AAC analysis of polarized parton distributions with uncertainties}

\author{M. Hirai$^*$, S. Kumano$^\dagger$, N. Saito$^\ddagger$ \\
                (Asymmetry Analysis Collaboration)}
\address{$^*$ Institute of Particle and Nuclear Studies, KEK \\ 
              1-1, Ooho, Tsukuba, 305-0801, Japan \\
         $^\dagger$ Department of Physics, Saga University,
                    Saga, 840-8502, Japan \\
         $^\ddagger$ Department of Physics, Kyoto University,
                    Kyoto, 606-8502, Japan}
                    
\maketitle

\abstracts{We report recent studies of the Asymmetry Analysis
    Collaboration (AAC) on polarized parton distribution functions (PDFs).
    Using the data on the spin symmetry $A_1$ in deep inelastic lepton 
    scattering, we investigate optimum polarized PDFs. Their uncertainties are 
    estimated by the Hessian method. The uncertainties are large for
    the polarized antiquark and gluon distributions. We discuss the role
    of accurate SLAC-E155 proton data on the determination of the PDFs.
    The obtained distributions are compared with other parametrization results.}
    
\vspace{-8.95cm}
{\hfill SAGA-HE-210-04}
\vspace{+8.05cm}

%%%%%%%%%%%%%%%%%%%%%%%%%%%%%%%%%%%%%%%%%%%%%%%%%%%%%%%%%%%%%%%%%%%%%%%%%%%%%%%%
%%%%%%%%%%%%%%%%%%%%%%%%%%%%%%%%%%%%%%%%%%%%%%%%%%%%%%%%%%%%%%%%%%%%%%%%%%%%%%%%
\section{Introduction}
\vspace{-0.1cm}    

Polarized parton distribution functions (PDFs) have been investigated
by using experimental data on the spin asymmetry $A_1$ in deep
inelastic lepton-nucleon scattering. Thanks to several experiments,
gross properties of the polarized PDFs are now known. However,
the nucleon spin structure is still far from complete understanding. 
It becomes clear that the fraction of nucleon spin carried by quarks
is small; however, the gluon contribution cannot be specified at this stage.
Although the nucleon spin is one of fundamental physics quantities,
we do not know how the spin consists of quark and gluon spins.

Because the polarized distributions cannot be completely fixed at this stage,
it is especially important to show their uncertainties for summarizing
the present knowledge and for planning future experimental projects. 
Such uncertainties have been investigated in the unpolarized PDFs
in the last several years, and there were also recent error analyses
for the polarized ones.  We report a new analysis by the Asymmetry Analysis
Collaboration (AAC) for the polarized PDFs and their uncertainties \cite{aac03}.

The AAC proposed the first parametrization in Ref. \cite{aac00}.
The additional data to the first AAC analysis (AAC00) are the SLAC-E155 proton
data. Because they are accurate, it is interesting to investigate their
effects on the determination of the polarized distributions. 
Namely, we discuss how the data reduce the uncertainties of the polarized PDFs.
Next, the global analysis was repeated by taking $\Delta g$=0 at $Q^2$=1 GeV$^2$
in order to show error correlation effects between the polarized antiquark
and gluon distributions.

This paper consists of the following.
In Sec. \ref{method}, a method is discussed for obtaining the polarized PDFs
and their uncertainties by analyzing $A_1$ data.
Analysis results are shown in Sec. \ref{results}, and they are summarized
in Sec. \ref{summary}.

%%%%%%%%%%%%%%%%%%%%%%%%%%%%%%%%%%%%%%%%%%%%%%%%%%%%%%%%%%%%%%%%%%%%%%%%%%%%%%%%
%%%%%%%%%%%%%%%%%%%%%%%%%%%%%%%%%%%%%%%%%%%%%%%%%%%%%%%%%%%%%%%%%%%%%%%%%%%%%%%%
\section{Analysis method}
\label{method}
\vspace{-0.1cm}  
                
The polarized PDFs are determined by analyzing the data of spin asymmetry
for inclusive lepton scattering in the deep inelastic region.            
The spin asymmetry $A_1$ is expressed by the structure functions,
$F_2$ and $g_1$, and the longitudinal-transverse ratio $R$:
\vspace{-0.05cm}
\begin{equation}
        A_1(x, Q^2)=\frac{g_1(x, Q^2)}{F_2(x, Q^2)}\,
               2 \, x \, [1+R(x, Q^2)] \, .
\end{equation}
The function $F_2$ is expressed in terms of unpolarized PDFs, for which we use
the GRV98 parametrization. For the function $R$, a SLAC analysis result
is used. The function $g_1$ is expressed in terms of the polarized
distributions:
\vspace{-0.05cm}
\begin{align}
g_1 (x, Q^2) = \frac{1}{2}\sum\limits_{i=1}^{n_f} e_{i}^2
   \bigg\{ & \Delta C_q(x,\alpha_s) \otimes [ \Delta q_{i} (x,Q^2)
    + \Delta \bar{q}_{i} (x,Q^2) ]    
\nonumber \\
     & + \Delta C_g(x,\alpha_s) \otimes \Delta g (x,Q^2) \bigg\},
\end{align}
where the symbol $\otimes$ indicates the convolution integral
$ f (x) \otimes g (x) = \int^{1}_{x} dy f(x/y) g(y) /y $. 
The functions $\Delta q$, $\Delta \bar{q}$, and $\Delta g$ are
polarized quark, antiquark, and gluon distributions.
The $\Delta C_q$ and $\Delta C_g$ are polarized coefficient functions
in the next-to-leading order.

The functional form of the polarized PDFs is chosen
at the initial $Q^2$ ($\equiv Q_0^2$): 
\begin{equation}
        \Delta f(x) = [\delta x^{\nu}-\kappa (x^{\nu}-x^{\mu})] f(x) \, ,
\label{eqn:df}
\end{equation}
where $f(x)$ is the unpolarized PDF. The factors $\delta$, $\nu$, $\kappa$,
and $\mu$ are the parameters to be determined by the $\chi^2$ analysis.
For $\Delta f$, we take $\Delta u_v$, $\Delta d_v$, $\Delta \bar q$,
and $\Delta g$ by assuming flavor symmetric antiquark distributions
($\Delta \bar u =\Delta \bar d=\Delta \bar s \equiv \Delta \bar q$)
at $Q_0^2$. The initial scale is taken $Q_0^2$=1 GeV$^2$.
The distributions are evolved to the experimental $Q^2$ points so that
the following $\chi^2$ is calculated:
\begin{equation}
\chi^2=\sum_i \frac{[ A_{1, \, i}^{\rm data}(x,Q^2)
                     -A_{1, \, i}^{\rm theo}(x,Q^2) ]^2}
                {[\Delta A_{1, \, i}^{\rm data}(x,Q^2) ]^2} ,
\end{equation}
where $A_{1, \, i}^{\rm data}$ and $A_{1, \, i}^{\rm theo}$ indicate
an asymmetry data and a calculated asymmetry, and
$\Delta A_{1, \, i}^{\rm data}$ is an experimental error.
The actual $\chi^2$ minimization is done by the CERN program library,
{\tt MINUIT}.
The data are taken by the EMC, SMC, SLAC-E130, E142, E143, E154, E155,
and HERMES collaborations. Recent JLab data will be included in 
our future analysis.

Uncertainties of the polarized PDFs are calculated by the Hessian method.
Running the {\tt MINUIT} subroutine, we obtain the Hessian matrix $H$,
which is then used for calculating the uncertainties:
\begin{equation}
        [\delta F(x)]^2=\Delta \chi^2 \sum_{i,j}
          \left( \frac{\partial F(x,a)}{\partial a_i}  \right)_{a=\hat a}
          H_{ij}^{-1}
          \left( \frac{\partial F(x,a)}{\partial a_j}  \right)_{a=\hat a} \ .
        \label{eq:erroe-M}
\end{equation}
Here, $F$ indicates a polarized PDF, $a_i$ is a parameter in the $\chi^2$ fit,
and $\hat a$ indicates the optimized parameter set. The $\Delta \chi^2$ value
determines the confidence level, and it is chosen as $\Delta \chi^2$=12.647
by considering that the number of parameters is eleven \cite{aac03}.
It corresponds to the one-$\sigma$-error range.

%%%%%%%%%%%%%%%%%%%%%%%%%%%%%%%%%%%%%%%%%%%%%%%%%%%%%%%%%%%%%%%%%%%%%%%%%%%%%%%%
%%%%%%%%%%%%%%%%%%%%%%%%%%%%%%%%%%%%%%%%%%%%%%%%%%%%%%%%%%%%%%%%%%%%%%%%%%%%%%%%
\section{Results}
\label{results}

\begin{figure}[h!]
\vspace{-0.3cm}
\begin{center}
     \includegraphics[width=0.400\textwidth]{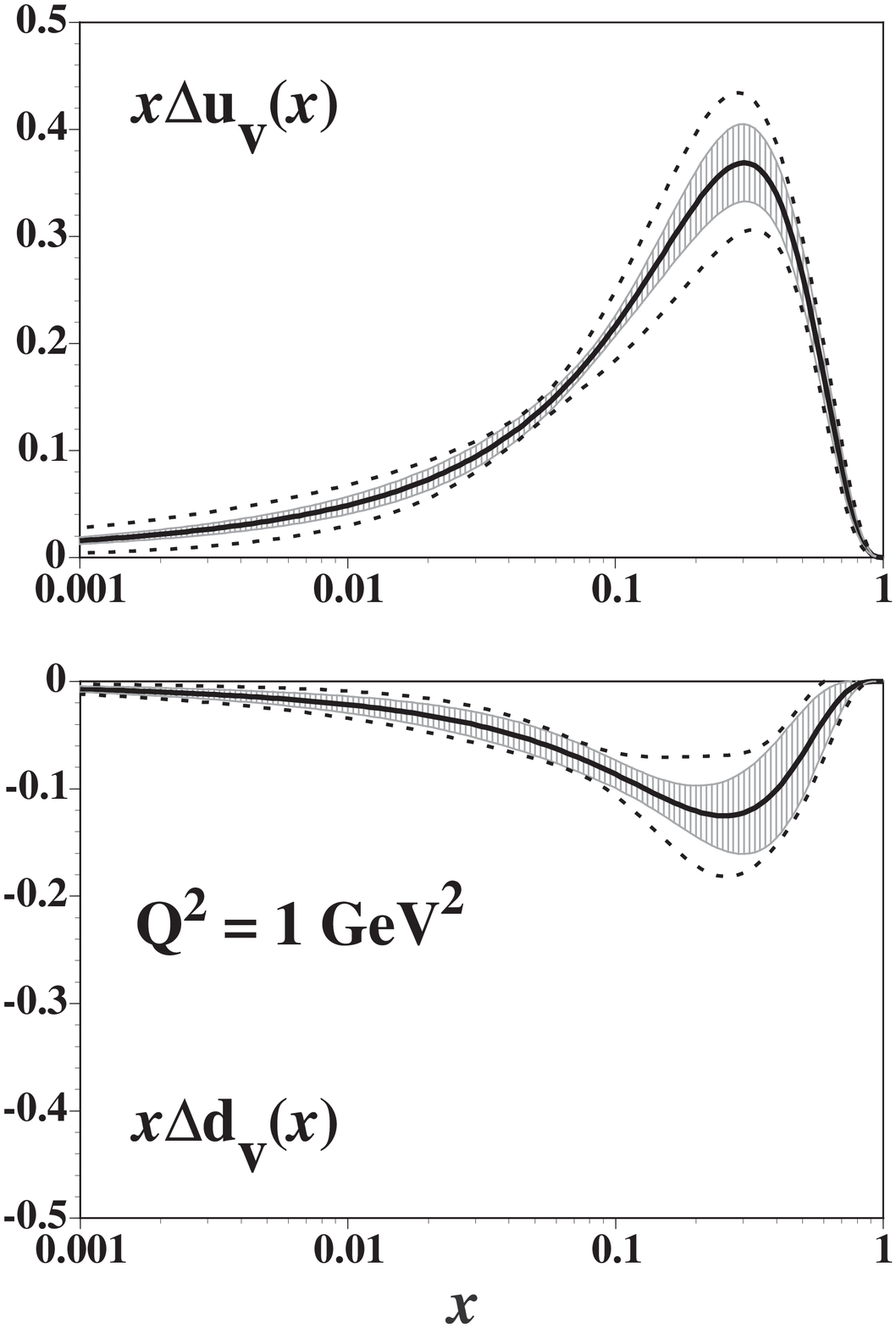}
\hspace{0.3cm}
     \includegraphics[width=0.407\textwidth]{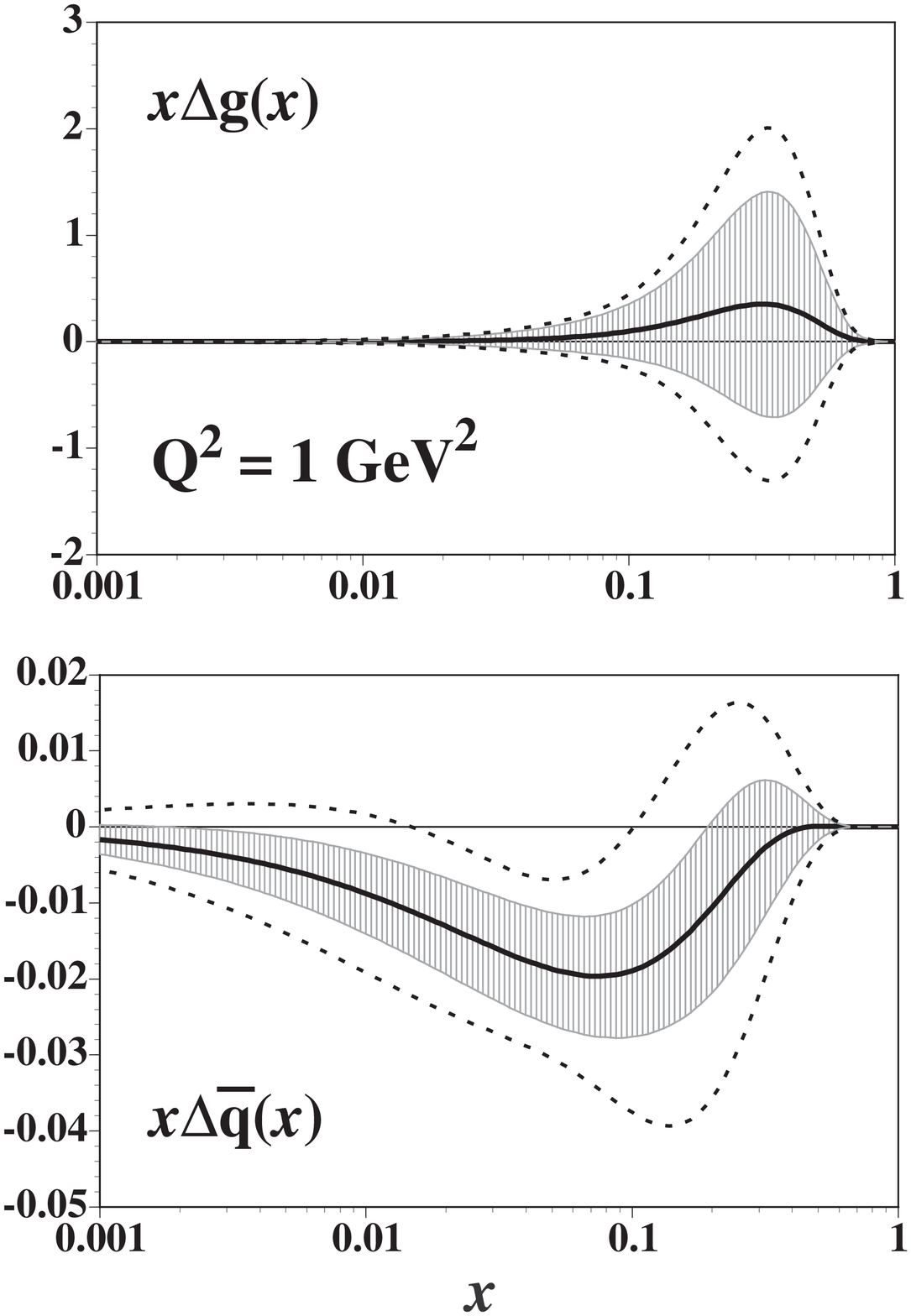}
\end{center}
\vspace{-0.2cm}
\caption{Polarized PDFs and their uncertainties are shown
         at $Q^2$=1 GeV$^2$. The solid curves and shaded areas indicate
         the AAC03 distributions and uncertainties. The dashed curves
         indicate the uncertainties of the previous version AAC00.}
\vspace{-0.5cm}
\label{fig:aac00-03}
\end{figure}

The optimum polarized PDFs obtained by the $\chi^2$ fit are shown in
Fig. \ref{fig:aac00-03}. The uncertainties are shown by the shaded
areas for the AAC03 version and by the dashed curves for
the previous version AAC00. The difference between AAC00 and AAC03 is
the addition of the accurate E155 proton data. It is obvious from
the figure that the accurate E155 data improves the PDF determination.
The magnitudes of the uncertainty bands are significantly reduced.
It is noteworthy that the gluon uncertainties are also reduced
in spite of the fact that the polarized gluon contribution to
$g_1$ is rather small. It should be caused by the error correlation
between the antiquark and gluon distributions.
In order to investigate such a correlation, we repeated the global 
analysis by terminating the gluon distribution ($\Delta g$=0)
at the initial $Q^2$ ($Q_0^2$).
Then, we find that the uncertainties of $\Delta \bar q$ are significantly
reduced, while those of $\Delta u_v$ and $\Delta d_v$ remain unchanged.
It suggests that the errors of the polarized antiquark and gluon
distributions should be strongly correlated. From these studies,
it is clear that an accurate determination of $\Delta g$ is important
for the determination of $\Delta \bar q$ and vice versa.

%%%%%%%%%%%%%%%%%%%%%%%%%%%%%%% figure %%%%%%%%%%%%%%%%%%%%%%%%%%%%%%%
%%%%%%%%%%%%%%%%%%%%%%%%%%%%%%% figure %%%%%%%%%%%%%%%%%%%%%%%%%%%%%%%

\begin{figure}[h]
\vspace{-0.0cm}
\begin{center}
     \includegraphics[width=0.400\textwidth]{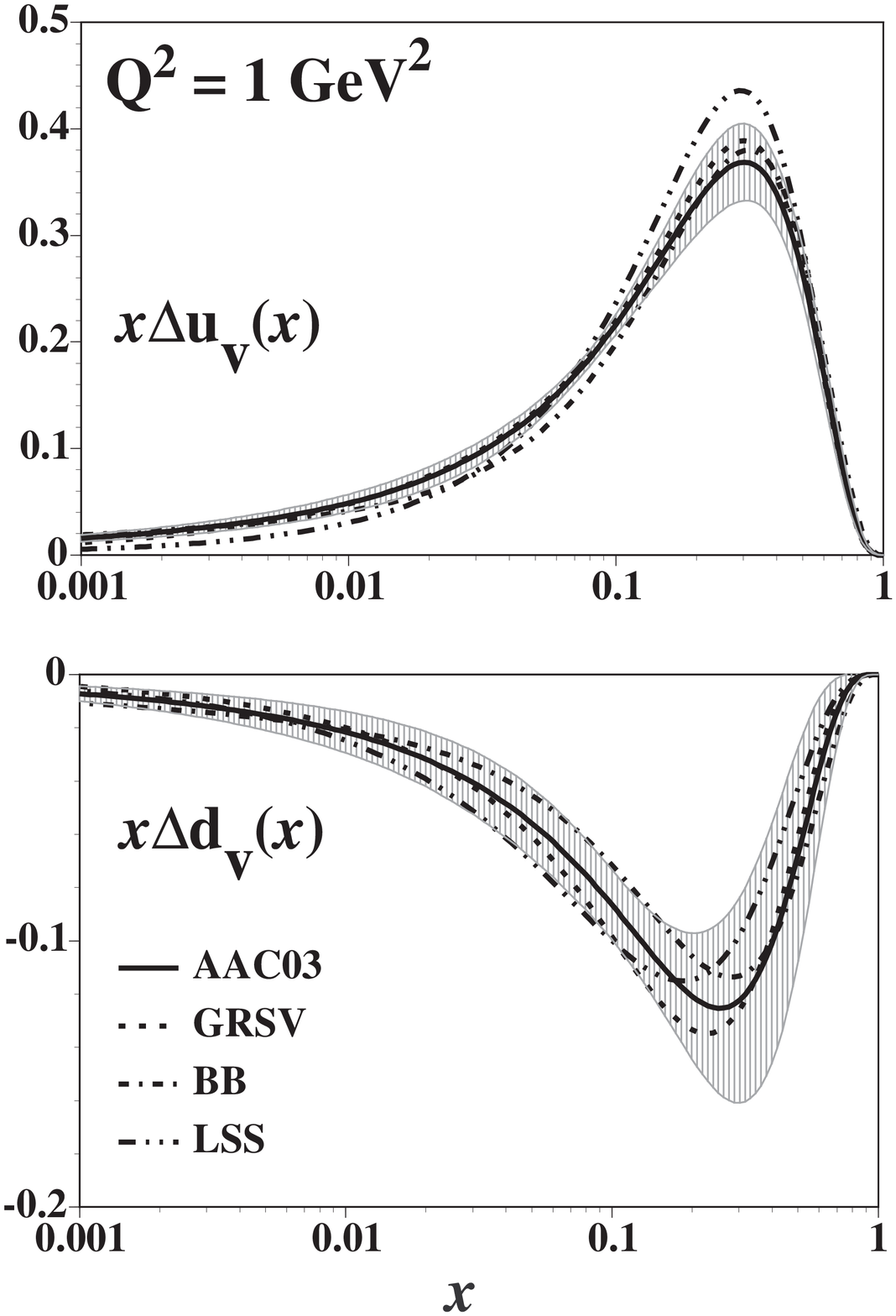}
\hspace{0.3cm}
     \includegraphics[width=0.407\textwidth]{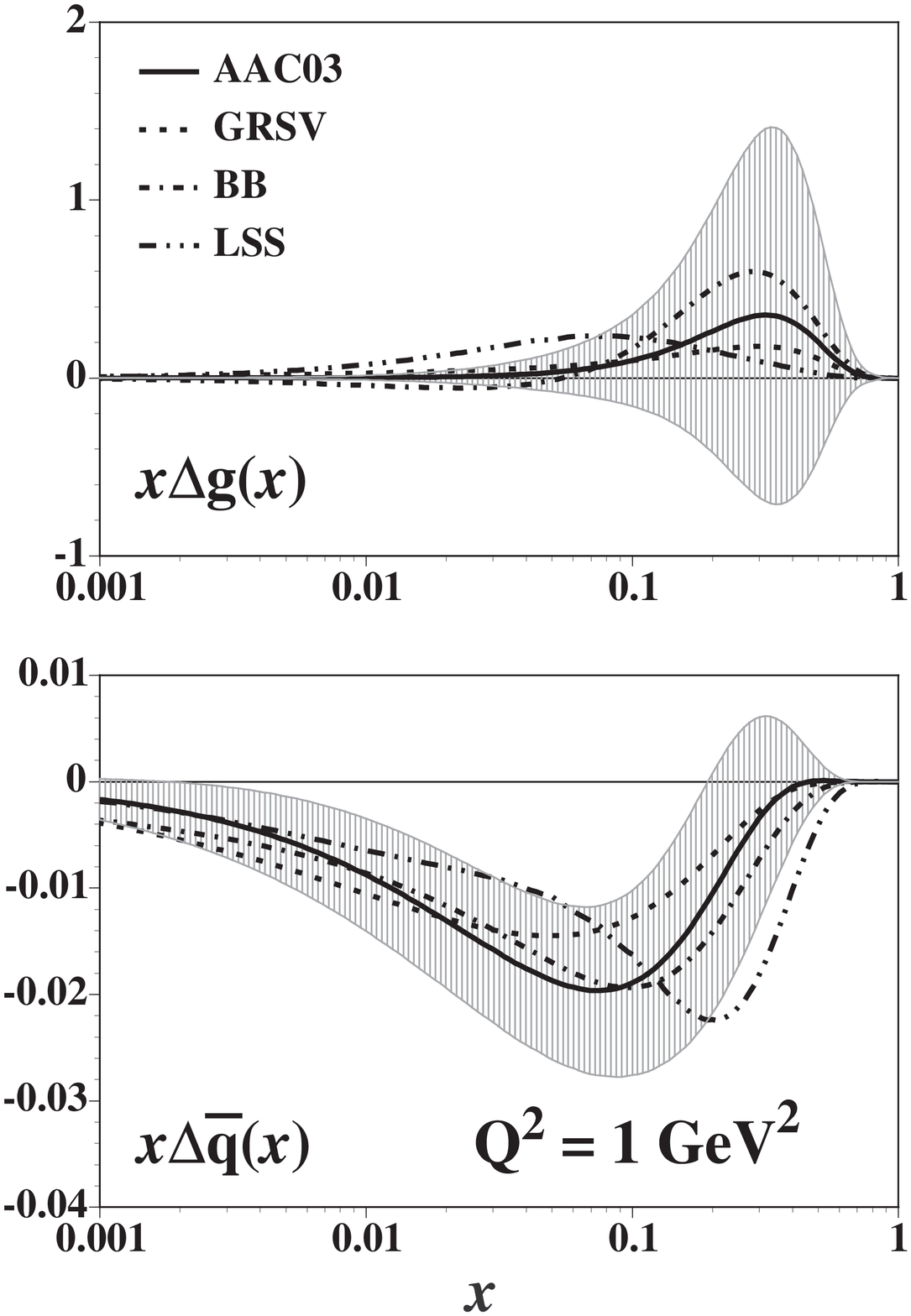}
\end{center}
\vspace{-0.2cm}
\caption{The polarized PDFs are compared with other parametrizations
         at $Q^2$=1 GeV$^2$.}
\vspace{-0.2cm}
\label{fig:comparison}
\end{figure}

The AAC03 distributions are compared with some of other analysis
results, GRSV \cite{grsv}, LSS \cite{lss}, and BB \cite{bb},
in Fig. \ref{fig:comparison}. All the distributions roughly agree,
and they are within the uncertainty bands.
The uncertainties of $\Delta g$ are especially large, which indicates
that even $\Delta g=0$ is allowed at this stage.
There are differences between the antiquark distributions
at small $x$, and it leads to different values of quark spin content
\cite{aac00}.

%%%%%%%%%%%%%%%%%%%%%%%%%%%%%%% figure %%%%%%%%%%%%%%%%%%%%%%%%%%%%%%%
%%%%%%%%%%%%%%%%%%%%%%%%%%%%%%% figure %%%%%%%%%%%%%%%%%%%%%%%%%%%%%%%

%%%%%%%%%%%%%%%%%%%%%%%%%%%%%%%%%%%%%%%%%%%%%%%%%%%%%%%%%%%%%%%%%%%%%%%%%%%%%%%%
%%%%%%%%%%%%%%%%%%%%%%%%%%%%%%%%%%%%%%%%%%%%%%%%%%%%%%%%%%%%%%%%%%%%%%%%%%%%%%%%
\section{Summary}
\label{summary}

We reported the AAC03 analysis for determining polarized parton
distribution functions. The uncertainties were estimated by the Hessian
method. Comparing the AAC03 results with the previous ones (AAC00), we
find that the E155 proton data played an important role in reducing
not only the uncertainties of $\Delta q$ and $\Delta \bar q$
but also the $\Delta g$ uncertainties due to the error correlation
between $\Delta \bar q$ and $\Delta g$. The uncertainties of $\Delta g$
are huge, so that we need future experimental measurements such as
prompt-photon measurements at RHIC \cite{mh}.

%%%%%%%%%%%%%%%%%%%%%%%%%%%%%%%%%%%%%%%%%%%%%%%%%%%%%%%%%%%%%%%%%%%%%%%%%%%%%%%%
\section*{Acknowledgements} 
\vspace{-0.1cm}
S.K. was supported by the Grant-in-Aid for Scientific Research from
the Japanese Ministry of Education, Culture, Sports, Science, and Technology. 

%%%%%%%%%%%%%%%%%%%%%%%%%%%%%%%%%%%%%%%%%%%%%%%%%%%%%%%%%%%%%%%%%%%%%%%%%%%%%%%%

%%%%%%%%%%%%%%%%%%%%%%%%%%%%%%%%%%%%%%%%%%%%%%%%%%%%%%%%%%%%%%%%%%%%%%%%%%%%%%%%

\end{document}